\documentstyle[aps,prl,floats]{revtex}
\begin{document}
\draft
\input epsf.sty

\twocolumn[\hsize\textwidth\columnwidth\hsize\csname
@twocolumnfalse\endcsname
\title{Spectroscopic Evidence
 for the Localization of Skyrmions near $\nu$=1 as $T\rightarrow 0$}
\author{P. Khandelwal$^{1}$,
A.\,E. Dementyev$^{1}$,
 N.\,N. Kuzma$^{1}$,
 S.\,E. Barrett$^{1}$,
L.\,N. Pfeiffer$^{2}$, and K.\,W. West$^{2}$}
\address{$^{1}$Department of Physics,
  Yale University, New Haven, Connecticut 06511
\\ $^{2}$Bell Laboratories, Lucent Technologies,
  Murray Hill, New Jersey 07974}
\date{\today}
\maketitle
\begin{abstract}
Optically pumped nuclear
magnetic resonance  
measurements of   $^{71}$Ga  spectra
were carried out in an n-doped GaAs/Al$_{0.1}$Ga$_{0.9}$As
multiple quantum well  sample near the integer
quantum Hall ground state $\nu$=1. As the temperature is lowered (down to $T \approx 0.3$ K),
a ``tilted plateau'' emerges in the Knight shift data,
which is a novel experimental signature of quasiparticle localization.
The dependence of the spectra on both
$T$ {\em and} $\nu$ suggests
that the localization is a collective process.  
The frozen limit spectra appear to rule out a 2D lattice of conventional Skyrmions.
\end{abstract}

\vskip 2pc ] 

\narrowtext

One of the most surprising twists in the recent history of
the quantum Hall effects\cite{QHEbook} was the prediction\cite{sondhi} that
novel spin textures called Skyrmions
can be the charged quasiparticles introduced by small deviations
($| \delta$$\nu |$) from ferromagnetic quantum Hall ground
states\cite{QHFerro}
(e.g., at Landau level filling factor $\nu$=$1 $ or $ \frac{1}{3}$).
A Skyrmion has an effective number of spin reversals $K$
and ``size" $\lambda$ that are determined by the competition between
the Coulomb energy (which increases both) and the
Zeeman energy (which reduces both).  Qualitatively, 
this cylindrically symmetric spin texture has a down spin at $r$=0
and a smooth radial transition to up spins at $r$=$\infty$.
In between, the nonzero XY spin components have a
vortical configuration\cite{sondhi,abolfathSkyrmion}.
The addition of Skyrmions to the
$\nu$=1 ground state was predicted to result in a rapid 
drop in the electron spin polarization, as $|\delta$$\nu|$ 
is increased\cite{fertigP(nu)}.
Several experiments are consistent with
this\cite{barrett,aifer,song} and other
predictions\cite{schmeller,tycko,bayot1and2,maude,melinte1}
of the Skyrmion model.  

These developments stimulated many
theoretical studies of Skyrmions.  One of the
central questions that emerged was the nature of the many-Skyrmion
ground state.  Would Skyrmions form a 
crystal\cite{brey,green,cote,rao,abolfathSkyrmeXtal,nazarovSkyrmeXtalQPT,TimmMelting},
and, if so, what symmetries would it possess?  
Does disorder\cite{nederveen} affect the
Skyrmion size $\lambda$ and spin number $K$ as $T\rightarrow 0$?
While these questions are still under active investigation,
basic aspects of the Skyrmion model
have not yet been tested experimentally.  For example,
the detailed shape of a Skyrmion has not yet been measured
with local probes, presumably because the quasiparticles 
are delocalized at high $T$. 

In this Letter, we report the first spectroscopic evidence
for Skyrmion localization.  The multiple quantum well sample used in this work was previously
studied at higher temperatures\cite{barrett,tycko}.  
The new data presented here were obtained by extending
the optically pumped nuclear
magnetic resonance (OPNMR) technique\cite{opnmr} to lower temperatures ($T\approx0.3$ K)
as described elsewhere\cite{khandelwal,kuzma}.

\begin{figure}
\centerline{\epsfxsize=2.9in\epsfbox{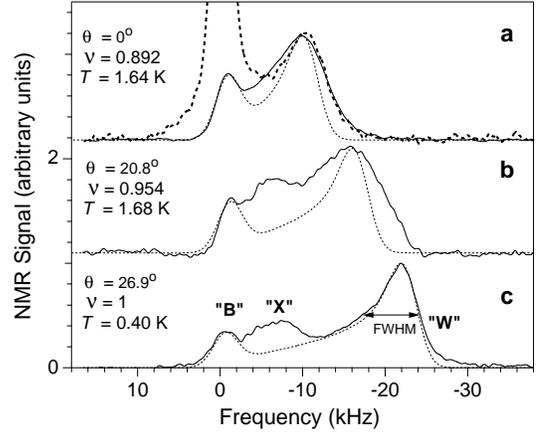}}
\caption{Several $^{71}$Ga OPNMR emission
spectra ({\bf a}-{\bf c}, solid lines, offset for clarity),
acquired at low $T$ for different $\nu$,
in $B_{\rm{tot}}$=$7.03$ T ($f_{o}$=91.36 MHz).  
The same ``$W$" resonance is obtained using
either conventional NMR absorption ({\bf a}, dashed line) or
OPNMR emission ({\bf a}, solid line).
The dotted lines ({\bf a}-{\bf c}) are fits  to $W$ and ``$B$", 
which assume identical quantum wells (see text). }
\label{fig1}
\end{figure}

Figure \ref{fig1} shows some OPNMR spectra
for $\nu$ close to one.  Nuclei within
the quantum wells are coupled to the spins of the
two-dimensional electron system via the isotropic Fermi contact
interaction\cite{slichter}, which shifts the
corresponding well resonance (labeled ``$W$" on
Fig.\,\ref{fig1}c)  relative to the signal from
the barriers (``$B$")\cite{barrett,kuzma}.
We define the Knight shift $K_S$ as the
peak-to-peak splitting between $W$
and $B$.  The third resonance 
(labeled ``$X$" in Fig.\,\ref{fig1}c) may
be due to nuclei near ``defect" sites.  By scaling equilibrium
signal intensities, we estimate that $X$ is due
to only $\sim 2\%$ of quantum well nuclei\cite{dementyevPre};
for now we ignore it.

For $\nu$=1, all spectra (e.g.,
Fig.\,\ref{fig1}c) are well described
by the same two-parameter fit (dotted lines)\cite{khandelwal,kuzma}
that was previously used for all spectra at $\nu$=$\frac{1}{3}$.
Note that this
fitting function has no explicit dependence on the
(x,y) position of nuclei along the quantum well, 
despite the fact that NMR is a local probe.  
This is because the fit is generated under the assumption
that all electron spins are {\em delocalized}, so that
$\langle S_{z}(\nu,T)\rangle$, averaged over the
NMR time scale ($\sim\,$20$\,\mu$sec), appears
spatially homogeneous along the plane of the wells,
and thus the resulting lineshape is
``motionally-narrowed'' \cite{slichter}.
In this limit, measurement of $K_S$ reveals the ``global",
time-averaged value of the electron spin polarization
${\cal P}$.  
On the other hand, for $\nu$$\not=$1 and low $T$, the well
resonance $W$ (Figs.\,\ref{fig1}a and \ref{fig1}b)  can also be
much broader than this same fit (dotted lines).

\begin{figure}
\centerline{\epsfxsize=3.7in\epsfbox{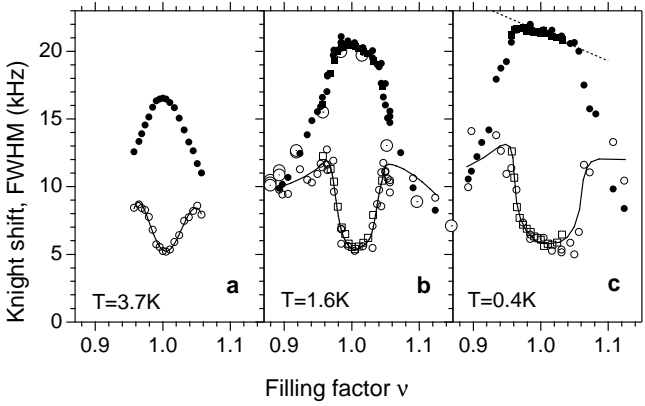}}
\caption{$K_{S}(\nu)$ (solid symbols)
 and $\Gamma_{w}$$(\nu)$ (small open symbols) at ({\bf a}) $T$=3.7 K,
 ({\bf b}) $T$=1.6 K, and ({\bf c}) $T$=0.4 K.  The large open symbols in ({\bf b})
are $K_S$ data previously reported for this sample.   
Here, $\Gamma_{w}$ is 
the full width at half maximum (FWHM) for the ``$W$'' resonance.
The filling factor ($\nu$=$nhc/(eB_{\rm{tot}}\!\cos\theta)$) is
varied {\em in situ} by tilting the sample 
($0^{\circ}$$\leq$$\theta$$\leq$$ 37^{\circ}$). In practice, all the low 
temperature (T$<$1.6 K) data were acquired following a cooldown from at least 1.6 K 
{\em at fixed} $\nu$.
The 2D electron density ($n=1.52\times10^{11}cm^{-2}$) is inferred from the $K_{S}(\nu)$ 
peak in ({\bf a}). 
Solid lines
are  to guide the eye, and  the dashed line is described in the text.}
\label{fig2}
\end{figure}

Figure \ref{fig2} shows the $K_{S}(\nu)$ and
the $\Gamma_{w}$$(\nu)$ of the $W$ resonance near $\nu = 1$,
for three different temperatures.  
As the temperature is lowered (Figs.\,\ref{fig2}a-\ref{fig2}c),
the sharp peak in $K_{S}(\nu)$ evolves into a ``tilted plateau".
Figure \ref{fig2}b also contains the $K_{S}(\nu)$ data points
reported previously\cite{barrett}.  While the new data are
consistent with the earlier measurements, probing $K_{S}(\nu)$
on this finer scale reveals a small region on both sides of $\nu$=$1$
where $K_{S}(\nu)$$\approx$$K_{S}(\nu$=$1)/\nu$ (dashed
line in Fig.\,\ref{fig2}c).  This tilted plateau
is incompatible with the expression for $K_{S}(\nu)$
derived previously\cite{barrett}, which had assumed
$\em{delocalized}$ $\em{quasiparticles}$\cite{formulaforp}. 
The existence of the 
tilted plateau  
is  a natural consequence of the localization of the
quasiparticles along
the plane of the quantum well, such that the nuclei responsible
for the $W$ resonance see fully polarized electrons ($\cal P$=1),
as if  $\nu$=1 ``locally", even though
$\nu$$\not=$1 ``globally".
  More precisely, for nuclei at ${\bf R_{i}}$=($X',Y',Z'$=0)
   in the center of the quantum well, 
  the 
local $K_{S\rm{int}}$(${\bf R_{i}}$)    
 is directly proportional
to the $z$-component of the local electron spin magnetization density,
$M_{z}$(${\bf R_{i}}$)\cite{slichter},
which is in turn proportional to the product of the 
electron number density and the spin polarization, i.e., 
 $M_{z}$(${\bf R_{i}}$)$\propto$
$|\phi$(${\bf R_{i}}$)$|^2$
$\cal P$(${\bf R_{i}}$)\cite{kuzma}.
If   the    quasiholes 
(or   quasiparticles) introduced into the system  by
 going to $\nu$\,=\,$1-\epsilon$ (or $\nu$\,=\,$1+\epsilon$) are localized, then,
in order to keep the total number of electrons fixed,
$|\phi$(${\bf R_{i}}$)$|^2$
{\em must} increase (or decrease) far from these charged excitations,
  which produces the 
observed tilt in the plateau near $\nu$=1.
  This is 
just the same effect as the rise (or fall)  of the water level in a pool
induced by placing solid dipsticks (or hollow capillary tubes) into it
to create localized density minima (or maxima), which correspond to
quasiholes  (or quasiparticles).
In this analogy, changing the filling
factor by adjusting $B_{\rm{tot}}\!\cos\theta$ varies  the number  
  of dipsticks  or capillary tubes in the pool and the Knight shift is given by the water level.
 Clearly,
the fact that NMR is a local probe has become important.

\begin{figure}
\centerline{\epsfxsize=3.5in\epsfbox{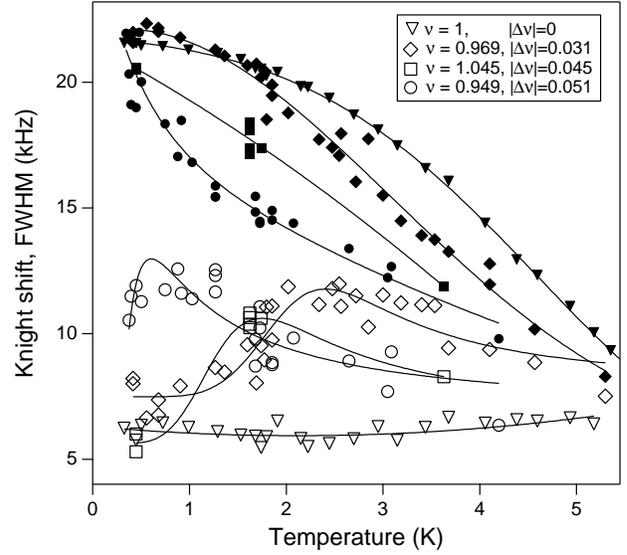}}
\caption{$K_S(T)$
(filled symbols) and $\Gamma_w(T)$ (open symbols) for
several filling factors 0.949$\leq$$\nu$$\leq$1.045.
 Lines through $K_S(T)$ are
to guide the eye. Lines through $\Gamma_w(T)$ are
fits suggested by a simple model for Skyrmion dynamics (see text).}
\label{fig3}
\end{figure}

Taking different slices through the ($\nu$, $T$) plane provides
additional insights.  Figure \ref{fig3} shows $K_{S}(T)$ and
$\Gamma_{w}$$(T)$ for several filling factors near $\nu$=1.
For $\nu$$\not=$1, lowering the temperature causes
$\Gamma_{w}$$(T)$ first to increase and then
to drop, in stark contrast to the
temperature independence of the well linewidth at $\nu$=1.
The non-monotonic temperature dependence (Fig.\ref{fig3})
is consistent with the evolution of the $W$ resonance 
from    motionally-narrowed  to   frozen      as the 
temperature is lowered.
Qualitatively similar trends were uncovered in
earlier measurements at $\nu$\,$<$\,$1/3$\cite{kuzma}.
Both cases are rather unusual examples of  
motional narrowing phenomena in NMR,
since the nuclei are fixed in the lattice at such low $T$.  
Instead, the motion is that of delocalized spin-reversed quasiparticles,
that results in 
fluctuations of the local hyperfine field 
$\delta   B^e_{z}({\bf R_{i}})   $ at each nuclear site ${\bf R_{i}}$.
The shape of the resonance is sensitive to  $\Theta({\bf R_{i}})$ $\equiv$
$\tau({\bf R_{i}})  \delta B^e_{z}({\bf R_{i}})^{i}\gamma$, where
$\tau({\bf R_{i}})$ is the characteristic time scale of
the fluctuations, and ${^{i}\gamma}$ is the nuclear 
gyromagnetic ratio\cite{slichter}.
As $T$ is lowered, the $W$ resonance goes from the 
motionally narrowed limit (at high $T$,
$\Theta({\bf R_{i}})<<1$)
to the ``intermediate limit" (at $T$ near $T_{max}$,
 where $\Gamma_{w}(T_{max})$=$\Gamma_{w}^{max}$, $\Theta({\bf R_{i}}) \sim 1$)
and then to the ``frozen limit" (at low $T$, 
$\Theta({\bf R_{i}})>>1$).    Figures \ref{fig2} and
\ref{fig3} show that all three limits are experimentally
accessible near $\nu$=1.

Qualitatively, the $|$$\delta$$\nu$$|$-dependence of
$\Gamma_{w}(T)$ (Fig. {\ref{fig3}})
appears to rule out several candidate mechanisms for the localization.
For example, if {\em individual} Skyrmions were strongly pinned
by a distribution of traps, then, as $\delta$$\nu$ increases,
quasiparticles should ``start'' to localize at
the same temperature, but ``finish'' at lower and lower temperatures. In contrast,
as $\delta$$\nu$ increases, the whole
$\Gamma_{w}(T)$ peak shifts to lower T (Fig. {\ref{fig3}}); e.g., 
at $T\approx 1.5$ K,
all Skyrmions can appear to be either localized 
(at $|$$\delta$$\nu$$|=0.031$)
or delocalized (at $|$$\delta$$\nu$$|=0.051$). We conclude that
a {\em collective} process is required to explain the trends 
in Figs.~\ref{fig2} and \ref{fig3}.  This process, however,
does not appear to be the ``melting" of a classical Skyrmion crystal,
since the classical melting
temperature should increase as the crystal density (and the bond energy)
increases.  Two other collective mechanisms are qualitatively consistent with
the data. In the first scenario, the data are explained by the 
thermally assisted melting of a
quantum Skyrmion crystal, which is approaching a quantum melting transition\cite{paredes}
 at some 
$|$$\delta$$\nu$$|>0.05$.  In the second, the delocalization is due to the
the``depinning" of a Skyrmion crystal, since the soft (stiff) bonds of
the crystal at low (high) $|$$\delta$$\nu$$|$ may easily (not easily)
stretch to match the disorder potential, resulting in a high (low)
depinning temperature\cite{fukuyama}.

Additional, quantitative information about the localization process\cite{dementyevPre} may be
obtained by simulating the effect of Skyrmion dynamics on the OPNMR
spectra\cite{sinova,dementyevPre}.  For example, to explain the
$\Gamma_{w}(T,|$$\delta$$\nu$$|=0.031)$ data, we require $K\approx3$ even as
$\Theta$ $\rightarrow$1 from below; the Skyrmion still has a ``large'' spin
even as it takes $\sim 20 \mu$s to travel over the inter-Skyrmion spacing.  Furthermore, 
in simulations that assume spatially-uniform local field fluctuations, we obtain
$\Gamma_{w}^{max}$ values that grow with increasing $|$$\delta$$\nu$$|$, 
as seen in earlier measurements\cite{kuzma}.  In contrast, Fig. {\ref{fig3}}
shows that $\Gamma_{w}^{max}$ is essentially constant for  $0.031<|$$\delta$$\nu$$|<0.051$.
Apparently, the actual localization process causes local field fluctuations that
become more inhomogeneous as $|$$\delta$$\nu$$|$ increases\cite{activation}.

In the frozen limit, the fit previously used for the
motionally narrowed OPNMR spectrum\cite{khandelwal,kuzma} is replaced by
the more general formula:

\begin{equation}     I(f)\!\!=
\!\!a_{\rm{b}}\,g(f)+\!\!\sum_{{\bf R_{i}}}^{Area}\!\!\int^{\textstyle ^{K_{S\rm{int}}({\bf R_{i}})}
}_{\textstyle _{0}}
\!\!\!\!\!\!\!\!\!\!\!\!\!\!\!\!\!\!\!\!\!\! df'  g(\,f\!-\!f'\,)\,
I_{\rm{w}}^{\rm{int}}(\textstyle
K_{S\rm{int}}({\bf R_{i}}),f')
\label{eq1}
\end{equation}

where  the sum runs over all
nuclei in  the center of the quantum well, and
$K_{S\rm{int}}$(${\bf R_{i}}$) $\propto$ $M_{z}({\bf R_{i}})$, which
in turn reflects the shape of individual 
quasiparticles and their spatial arrangement along    
the plane of the quantum well.  Since conventional Skyrmions
are expected to form a square lattice, we consider this possibility
first.
  
\begin{figure}
\centerline{\epsfxsize=3.4in\epsfbox{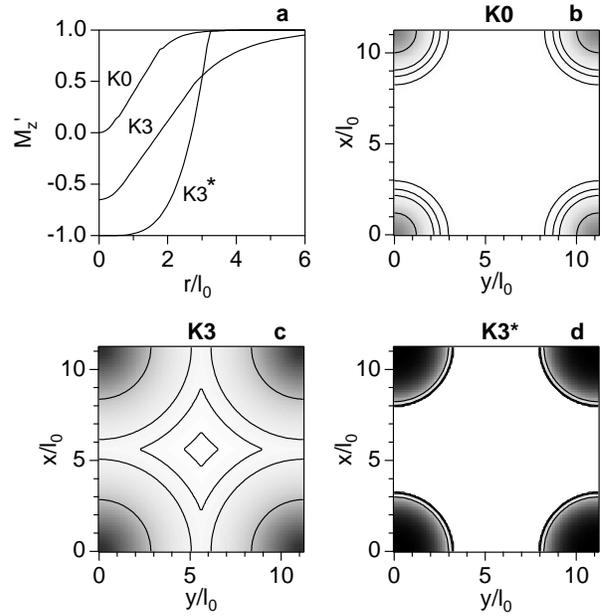}}
\caption{({\bf a}) Expected radial dependence (in units of the 
magnetic length $l_o$) of  $M'_{z}$($r$),
for the K0, K3, and K3$^{\ast}$ Skyrmions described in the text.  
For $|$$\delta$$\nu$$|$=$0.05$, 
gray scale images of $M_{z}^{Sim}$(${\bf R_{i}}$) (Black=$-$1; White=+1)  are shown 
within the unit cell of a square lattice of either
   ({\bf b}) K0,  ({\bf c}) K3, or  ({\bf d}) K3$^{\ast}$ Skyrmions.
Also shown (in {\bf b}-{\bf d}) are black contour lines 
at $M_{z}^{Sim}$= 0.5, 0.9, 0.95, and 0.98.}
\label{fig4}
\end{figure}

Several theoretical approaches have been used to calculate
  $M_{z}$(${\bf R_{i}}$) for
a  {\em single} Skyrmion excitation of the 
$\nu=1$ ground state.
  Figure {\ref{fig4}}a
shows the typical radial dependence of the dimensionless $M'_{z}$($r$)  
expected\cite{abolfathSkyrmion}
 for   Skyrmions with $K$=0    (``K0'') and  
     $K$=3    (``K3'') reversed spins. 
The K0  Skyrmion corresponds to 
  the ordinary Laughlin  quasiparticle, 
while both theory and experiment suggest
that the K3 Skyrmion is energetically
preferred for typical experimental conditions.
Also shown in Fig. {\ref{fig4}}a is an 
 ad hoc  hybrid  between the two (``K3$^{\ast}$''), that 
has both the tail of K0 and the three reversed spins of K3. 

\begin{figure}
\centerline{\epsfxsize=3.4in\epsfbox{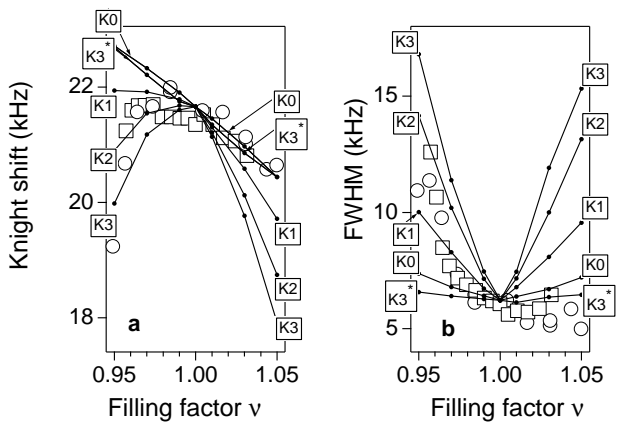}}
\caption{ Open symbols are [$K_{S}$$(\nu)$,  
$\Gamma_{w}$$(\nu)$] data  from Fig. 2{\bf c}.  Filled points are 
[$K_{S}^{Sim}$$(\nu)$,  
$\Gamma_{w}^{Sim}$$(\nu)$]  extracted from simulations
described in the text. The points for each Skyrmion type
are joined by lines.}
\label{fig5}
\end{figure}

We approximate $M_{z}$(${\bf R_{i}}$) for N localized Skyrmions
with $M_{z}^{Sim}$(${\bf R_{i}}$) (Fig. \ref{fig4}b-\ref{fig4}d).
Using Eqn. (\ref{eq1}), we simulate the  frozen limit OPNMR spectrum
at   $\nu$=$1+\delta\nu$, for various Skyrmion shapes.
Figure {\ref{fig5}} shows
[$K_{S}^{Sim}$$(\nu)$, $\Gamma_{w}^{Sim}$$(\nu)$] 
extracted from these simulations,
which may be quantitatively compared to the low temperature
data of  Fig. {\ref{fig2}}c.
We can {\em rule out} the standard model of a square lattice
of conventional K3 Skyrmions over most of the plateau
(see Fig. \ref{fig5}a and \ref{fig5}b); using smaller conventional
Skyrmions (K2, K1) helps, but not enough.  Instead, the data are
in much better agreement with simulations assuming a square lattice
of either K0 or ad-hoc K3$^{\ast}$ Skyrmions.  Apparently,
the existence of the tilted plateau requires that 
$M_{z}^{Sim}$(${\bf R_{i}}$) $\sim$ 1
over a large fraction of the area between the quasiparticles,
as in Figs. \ref{fig4}b and \ref{fig4}d.
This conclusion is not sensitive to the details of
our simulation over the range 
($|$$\delta$$\nu$$|\leq0.05$) of the observed plateau (e.g., changing
to a triangular lattice, or including small disorder in Skyrmion
locations). If the localized state
is a 2D lattice of quasiparticles, it appears that
they are either Laughlin quasiparticles (K0) or Skyrmions with very
short tails, like the ad-hoc K3$^{\ast}$.  

Alternatively, the localized state may involve ``clumps'' of Skyrmions,
which result in large Skyrmion-free regions. This may happen, for example,
if the disorder potential favors large length scale
density fluctuations, subject to the constraint of a collective
localization process (see Figs. \ref{fig2} and \ref{fig3}).
More theoretical work is required before this picture can be compared to
the data.  

For a 2D lattice, the Skyrmion shape consistent with
the data is surprising, since the energetics
of a {\em single Skyrmion} state would favor the 
conventional K3 over a shorter tailed Skyrmion (like
K3$^{\ast}$). However, this preference may not 
be the same in a {\em many Skyrmion} state, 
where  the energetics  are more complicated.
 For example, the energy of a single Skyrmion
is independent of the phase angle $\phi$ which defines
    the global orientation of the XY spin components.
In the crystalline phase, however, interactions 
are generally expected to lead to preferred values
for the relative 
phase angle ($\phi_{i}$-$\phi_{j}$) between the Skyrmions at  sites
($i,j$)\cite{brey}. 
Recent theoretical studies of a square lattice of K3 Skyrmions suggested
that it was equivalent to the superconducting phase (SC) of the 
boson Hubbard model\cite{cote}. 
We speculate that the shorter tail  of the K3$^{\ast}$ Skyrmion
would allow  $\phi_{i}$,$\phi_{j}$ to remain uncorrelated, thereby fixing
the number
of reversed spins on each site $K_{i}$ to an integer, which corresponds to
the Mott insulating phase (MI) of the 
boson Hubbard model\cite{nazarovSkyrmeXtalQPT}.  If the total energy
of this MI phase is lower than the SC
phase, the system prefers the K3$^{\ast}$ Skyrmion.

We thank J. Sinova, S. M. Girvin, A. H. MacDonald,
     H. A. Fertig, S. L. Sondhi,
     N. Read, S. Sachdev, and R. Shankar
     for many helpful
     discussions.   This work was
     supported by NSF Grant $\#$DMR-9807184.
     SEB also acknowledges an Alfred P.  Sloan
     Research Fellowship.


\begin{references}
\bibitem{activation}If we assume that $\tau({\bf R_{i}})$ follows an Arrhennius law,
					with a spread ($\Delta U$) about the average 
					activation energy ($U$), we can fit (Fig. {\ref{fig3}}, solid lines)
					the $\Gamma_{w}(T)$ data; $\Delta U$/$U$ grows from $\sim$1/8 to $\sim$1 
					as $|$$\delta$$\nu$$|$ increases\cite{dementyevPre}.
\end{references}
  \end{document}